\documentclass[12pt,draft]{article}

\usepackage{amssymb}
\usepackage{latexsym}

\newtheorem{theorem}{Theorem}[section]
\newtheorem{proposition}[theorem]{Proposition}

\newcommand{\nc}{\newcommand}
\nc{\C}{{\mathbb C}}
\nc{\R}{{\mathbb R}}
\nc{\HH}{{\mathbb H}}
\nc{\Z}{{\mathbb Z}}
\nc{\dd}{{\rm d}}
\nc{\ii}{{\bf i}}
\nc{\jj}{{\bf j}}
\nc{\kk}{{\bf k}}
\nc{\ad}{\mathop{\rm ad}\nolimits}
\nc{\tr}{\mathop{\rm tr}\nolimits}
\nc{\su}{{\mathfrak s}{\mathfrak u}(2)}
\nc{\so}{{\mathfrak s}{\mathfrak o} (4)}

\begin{document} 

\title{Geometric construction of new Yang--Mills instantons over Taub-NUT 
space} 

\author{
G\'abor Etesi
\\ {\it Yukawa Institute for Theoretical Physics,}
\\ {\it  Kyoto University,}
\\{\it Kyoto 606-8502, Japan}
\\ {\tt etesi@yukawa.kyoto-u.ac.jp}
\\
\\ Tam\'as Hausel
\\ {\it Miller Institute for Basic Research in Science and}\\
{\it Department of Mathematics,}
\\ {\it University of California at Berkeley,}
\\ {\it Berkeley CA 94720, USA}\\
{\tt hausel@math.berkeley.edu} }

\maketitle

\pagestyle{myheadings}
\markright{G. Etesi, T. Hausel: Constructing new Taub-NUT instantons}  

\thispagestyle{empty}

\begin{abstract}
In this paper we exhibit a one-parameter family of new Taub-NUT instantons
parameterized by a half-line.  The endpoint of the
half-line will be the reducible Yang--Mills instanton corresponding to the
Eguchi--Hanson--Gibbons $L^2$ harmonic
$2$-form, while at an inner point we recover the Pope--Yuille 
instanton constructed as a projection of the Levi--Civit\'a
connection onto the positive $\su^+\subset \so$ subalgebra. Our method
imitates the Jackiw--Nohl--Rebbi construction originally designed for
flat $\R^4$. That is we find a one-parameter family of harmonic functions
on the Taub-NUT space with a point singularity, rescale the metric and
project the obtained Levi--Civit\'a connection onto the other negative
$\su^-\subset\so$ part. Our solutions will possess the full $U(2)$
symmetry, and thus provide more solutions to the recently proposed $U(2)$
symmetric ansatz of Kim and Yoon. 

\end{abstract}

\newpage

\section{Introduction}
Motivated by Sen's S-duality conjecture \cite{sen} from 1994,  
recently there have been some interest in understanding the $L^2$-cohomology of
certain hyper-K\"ahler moduli spaces of magnetic monopoles. Probably the 
strongest evidence in Sen's paper for his conjecture was
an explicit construction of an $L^2$ harmonic $2$-form on the universal 
double cover of the Atiyah--Hitchin manifold. Sen's conjecture also
predicted an $L^2$ harmonic $2$-form on the Euclidean Taub-NUT space (in
what follows we call this space systematically the Taub-NUT
space). This was found later by Gibbons in 1996 \cite{gib}. He constructed
it as the exterior derivative of the dual of the Killing field of a
canonical $U(1)$-action. 

In \cite{ete-hau} we imitated Gibbons' construction and found one self-dual
(and one anti-self-dual) $L^2$ harmonic $2$-form on the Euclidean 
Schwarzschild manifold. We then went on and interpreted such a $2$-form
with integer cohomology class, as the curvature of a $U(1)$ Maxwell
connection. (We note that in this Maxwell form our solution was written 
down by Pope in \cite{pop} while the above Gibbons' $2$-form
was discovered by Eguchi and Hanson \cite{egu-han}, both in
1978.) This way we found reducible $SU(2)$ Yang--Mills connections on the
Euclidean Schwarzschild manifold, and showed that they agree with $SU(2)$
Yang--Mills connections found by Charap and Duff in 1978 \cite{cha-duf}
and were supposed to be irreducible solutions. 

In this paper we can start by interpreting the above
Eguchi--Hanson--Gibbons $L^2$ harmonic $2$-form as a reducible $SU(2)$
Yang--Mills instanton on the Taub-NUT space. 
We will show how to deform the one with unit energy to get a
one-dimensional family of $SU(2)$ Yang-Mills instantons on Taub-NUT space. 

There are only very few $SU(2)$ Yang--Mills instantons on Taub-NUT space
found in the literature compared to the case of flat $\R^4$ (see
\cite{jac-noh-reb} and \cite{ati-dri-hit-man}) and
in general to ALE spaces (see \cite{kro-nak} and also \cite{bia-fuc-ros-mar}).
There is a complete ADHM-Nahm data for the
$U(1)$-invariant self-dual configuration on Taub-NUT space described in
\cite{che-kap}.  
However apart from the above reducible ones apparently only the Pope--Yuille
instanton \cite{pop-yui} is known explicitly, which is of unit energy. 
In fact Pope and Yuille constructed it as the projection of the
Levi--Civit\'a connection onto the positive $\su^+\subset\so$ part (see
next section for details). 

The Pope-Yuille solution was reinvented by Kim and Yoon recently \cite{kim-yoo}. They
wrote down the full $U(2)$ symmetric ansatz, and came up with two coupled
differential equations. They were able to find a solution of unit energy
to this system, which turns out to be identical with the
above mentioned Pope--Yuille solution.

In the present paper we will use the method of Jackiw, Nohl and Rebbi
\cite{jac-noh-reb} to construct a one-dimensional family of $SU(2)$
Yang--Mills instantons of unit energy on Taub-NUT space. Namely we find
all $U(2)$-invariant harmonic functions on Taub-NUT space. They will be of
the form 
\[f(r)=1+{\lambda\over r-m}\]
for positive $\lambda$'s and will have a one-point singularity at the NUT. 
We produce our one-parameter (i.e. $\lambda\in (0,\infty]$) family of
instantons (see (\ref{megoldas}) for an explicit form) via  
rescaling the metric by $f$ and then projecting the
obtained Levi--Civit\'a connection onto the negative $\su^-\subset\so$
subalgebra, and finally removing the singularity at the NUT by applying
Uhlenbeck's theorem \cite{uhl}. 

When $\lambda =2m$ we recover the Pope--Yuille solution. When
$\lambda=\infty$, that is, after rescaling the metric with the harmonic 
function $1/(r-m)$, we will find a
reducible connection, namely the Eguchi--Hanson--Gibbons $L^2$ harmonic
form of unit energy. Thus we can interpret this result as a new
intrinsically geometric construction of this harmonic $2$-form. Also as 
a check we show that the explicitly calculated $SU(2)$-instantons are 
satisfying the Kim--Yoon ansatz \cite{kim-yoo}. 

Finally we note that the case of the multi-centered metrics of 
Gibbons--Hawking 
(cf. \cite{gib-haw}) will be treated  elsewhere \cite{ete-hau3}.
 
\vskip.4cm 

\paragraph{\bf Acknowledgement.} The first steps in this work were done
when the first author visited the University of California at Berkeley
in February 2001. We would like to acknowledge the financial support
provided by the Miller Institute of Basic Research in Science, the Japan
Society for the Promotion of Science, grant No. P99736 and the partial
support by OTKA grant No. T032478. 
 
G. E. is also grateful to Dr. A. Ishibashi (Yukawa Institute) for calling
his attention to Garfinkle's paper.

\section{The Atiyah--Hitchin--Singer theorem}

First we recall the general theory from \cite{boo-ble}.
Let $(M,g)$ be a four-dimensional Riemannian spin-manifold. Remember that
via Spin$(4)\cong SU(2)\times SU(2)$ we have a Lie algebra isomorphism
$\so\cong\su^+\oplus\su^-$. Consider the Levi--Civit\'a connection
which is locally represented by an $\so$-valued 1-form $\omega$ on
$TM$. Because $M$ is spin and four-dimensional, we can consistently lift
this connection to the spin connection, locally given by $\omega_S$, on
the spin bundle $SM$ (which is a complex bundle of rank four) and can
project it to the $\su^\pm$ components. The projected connections $A^\pm$
live on the chiral spinor bundles $S^\pm M$ where the decomposition
$SM=S^+M\oplus S^-M$ corresponds to the above splitting of Spin$(4)$. One
can raise the question what are the conditions on the metric $g$ for
either $A^+$ or $A^-$ to be self-dual (seeking for antiself-dual solutions
is only a matter of reversing the orientation of $M$).

Consider the curvature 2-form $R\in C^\infty (\Lambda^2M\otimes\so )$ of
the metric. There is a standard linear isomorphism
$\so\cong\Lambda^2\R^4$ given by $A\mapsto\alpha$ with $xAy=\alpha
(x,y)$ for all $x,y\in\R^4$. Therefore we may regard $R$ as a {\it
2-form-valued 2-form}
in $C^\infty (\Lambda^2M\otimes \Lambda^2M)$ i.e. for vector fields $X,Y$
over $M$ we have $R(X, Y)\in C^\infty (\Lambda^2M)$. Since the space
of four dimensional curvature tensors, acted on by $SO(4)$, is 20
dimensional, one gets a 20 dimensional
reducible representation of $SO(4)$ (and of Spin$(4)$, being $M$
spin). The decomposition into irreducible components is (see \cite{bes},
pp. 45-52 or \cite{boo-ble}, pp. 344-348)
\begin{equation}
R={1\over 12}\pmatrix{s & 0\cr 0 & s\cr}+\pmatrix{0 &
B\cr B^T & 0\cr}+
\pmatrix{W^+ & 0\cr 0 & 0\cr}+\pmatrix{0 & 0\cr 0 & W^-\cr},
\label{gorbuletszetszedes}
\end{equation}
where $s$ is the scalar curvature, $B$ is the traceless Ricci tensor,
$W^\pm$ are the Weyl tensors. The splitting of the Weyl tensor is a
special four-dimensional phenomenon and is related with the above 
splitting of the Lie algebra $\so$. There are two Hodge operations which
can operate on $R$. One (denoted by $*$) acts on the {\it 2-form part
of $R$} while the other one (denoted by $\star$) acts on the {\it values}
of $R$ (which are also 2-forms). In a local coordinate system, these
actions are given by
\[(\star R)_{ijkl}={1\over
2}\sqrt{\det g}\:\varepsilon_{ijmn}R^{mn}_{\:\:\:\:\:\:kl},\]
\[(*R)_{ijkl}={1\over
2}\sqrt{\det g}\:R_{ij}^{\:\:\:\:mn}\varepsilon_{mnkl}.\]
It is not difficult to see that the projections $p^\pm
:\so\rightarrow\su^\pm$ are given by $R\mapsto F^\pm:={1\over 2}(1\pm\star
R)$, and $F^\pm$ are self-dual with respect to $g$ if
and only if $*(1\pm\star R)=(1\pm\star R)$. Using the
previous representation for the decomposition of $R$ suppose $\star$ acts
on the left while $*$ on the right, both of them via
  \[\pmatrix{{\rm id} & 0\cr 0 & {\rm -id}\cr}.\]
In this case the previous self-duality condition looks like
($\overline{W}^\pm :=W^\pm+{1\over 12}s$)
\[\pmatrix{\overline{W}^+\pm\overline{W}^+ & -(B\pm B)\cr B^T\mp B^T&
-\left(\overline{W}^-\mp\overline{W}^-\right)\cr}=   
\pmatrix{\overline{W}^+\pm\overline{W}^+
&B\pm B\cr B^T\mp B^T & \overline{W}^-\mp\overline{W}^-\cr}.\]
From here we can immediately conclude that $F^+$ is self-dual if and
only if $B=0$ i.e. $g$ is {\it Einstein} while $F^-$ is self-dual if
and only if $\overline{W}^-=0$ i.e. $g$ is {\it
half-conformally flat (i.e. self-dual) with vanishing scalar
curvature}. Hence we have proved \cite{ati-hit-sin}:
\begin{theorem} [Atiyah--Hitchin--Singer] Let
$(M,g)$ be a four-dimensional Riemannian spin manifold. Then 

\noindent {\rm (i)} $F^+$ is the curvature of
an self-dual $SU(2)$-connection on $S^+M$ if and only if $g$ is Einstein,
or

\noindent {\rm (ii)} $F^-$ is the curvature of a self-dual
$SU(2)$-connection on $S^-M$ if and only if $g$ is half
conformally flat (i.e. self-dual) with vanishing scalar
curvature. $\Diamond$
\end{theorem}
Remember that both the (anti)self-duality equations
\[*F=\pm F\]
and the action
\[\Vert F\Vert^2_{L^2(M)}={1\over 8\pi^2}\int\limits_M\vert
F\vert^2_g=-{1\over 8\pi^2}\int\limits_M\tr (F\wedge *F)\]
are conformally invariant in four dimensions; consequently if we can
rescale $g$ with a suitable positive function $f$ producing a metric
$\tilde g$ which satisfies one of the properties of the previous theorem
then we can construct instantons over the original manifold $(M,g)$. This
idea was used by Jackiw, Nohl and Rebbi to construct instantons over the
flat $\R^4$ \cite{jac-noh-reb}. 

\section{Projecting onto the positive side}

First consider the case of $F^+$, i.e. part (i) of the above theorem. Let 
$(M,g)$ be a Riemannian manifold of dimension $n$. Remember that $\psi
:M\rightarrow M$ is a {\it conformal isometry} of
$(M,g)$ if there is a function $f: M\rightarrow\R$ such that
$\psi^*g=f^2g$. Notice that being $\psi$ a diffeomorphism, $f$ cannot be
zero anywhere i.e. we may assume that it is positive, $f>0$. Ordinary
isometries are the special cases with $f=1$. The
vector field $X$ on $M$, induced by the conformal isometry, is
called a {\it conformal Killing field}. It satisfies the {\it conformal
Killing equation} (\cite{wal}, pp. 443-444) 
\[L_Xg-{2{\rm div}(X)\over n}g=0\]
where $L$ is the Lie derivative while div is the divergence of a
vector field. If $\xi =\langle X,\:\cdot\:\rangle$ denotes the dual 1-form
to $X$ with respect to the metric, then consider the following {\it
conformal Killing data}:
\begin{equation}
(\xi ,\:\:\:\dd\xi ,\:\:\:{\rm div}(X),\:\:\: \dd {\rm div}(X)).
\label{adatok}
\end{equation}
These satisfy the following equations (see \cite{gar} or \cite{ger}):
\begin{equation}
\begin{array}{ll}
\nabla\xi =(1/2)\dd\xi +(1/n){\rm div}(X)g,\\
\\
\nabla (\dd\xi )=(1/n)(g\otimes\dd {\rm div}(X) -\dd{\rm 
div}(X)\otimes g)+2R(\:\cdot\:,\:\cdot\:,\:\cdot\:,\xi ),\\
\\
\nabla ({\rm div}(X))=\dd{\rm div}(X),\\
\\
\nabla (\dd {\rm div}(X))=-(n/2)\nabla_XP-{\rm
div}(X)P-(n/2)\left\langle g\:,\: (P\otimes\dd\xi +\dd\xi\otimes
P)\right\rangle .\\
\end{array}
\label{egyenletek}
\end{equation}
Here $R$ is understood as the $(3,1)$-curvature tensor while $P=r-{1\over
6}sg$ with $r$ being the Ricci-tensor.
If $\gamma$ is a smooth curve in $M$ then fixing conformal Killing data in
a point $p=\gamma (t)$ we can integrate (\ref{egyenletek}) to get all the
values of $X$ along $\gamma$. Actually if $X$ is a conformal Killing field
then by fixing the above data in one point $p\in M$ we can determine the
values of $X$ over the {\it whole} $M$, provided that $M$ is connected. 
Consequently, if these data vanish in one point, then $X$ vanishes over
all the $M$. 

Furthermore a Riemannian manifold $(M,g)$ is called {\it irreducible} if 
the holonomy group, induced by the metric, acts irreducibly on each
tangent space of $M$. 

Now we can state:

\begin{proposition} 
Let $(M,g)$ be a connected, irreducible, Ricci-flat Riemannian
manifold of dimension $n>2$. Then $(M,\tilde{g})$ with
$\tilde{g}=\varphi^{-2}g$ is Einstein if and only if $\varphi$ is a
non-zero constant function on $M$. 
\end{proposition}

\noindent {\it Remark.} Notice that the above proposition is not true
for reducible manifolds: the already mentioned Jackiw--Nohl--Rebbi
construction \cite{jac-noh-reb} provides us with non-trivial Einstein metrics,
conformally equivalent to the flat $\R^4$. 
\vspace{0.1in}

\noindent {\it Proof}. If $\eta$ is a 1-form on $(M,g)$ with
dual vector $Y=\langle\eta ,\:\cdot\:\rangle$, then the $(0,2)$-tensor
$\nabla\eta$ can be decomposed into antisymmetric, trace and traceless
symmetric parts respectively as follows (e.g. \cite{pet}, pp. 200,
Ex. 5.):
\begin{equation}
\nabla\eta={1\over 2}\dd\eta -{\delta\eta\over n}g+{1\over 2}\left(
L_Yg+{2\delta\eta\over n}g\right)
\label{dekompozicio}
\end{equation}
where $\delta$ is the exterior codifferentiation on $(M,g)$
satisfying $\delta\eta =-{\rm div}(Y)$. Being $g$ an Einstein metric, it
has identically zero traceless Ricci tensor i.e. $B=0$ from the
decomposition (\ref{gorbuletszetszedes}). We rescale
$g$ with the function $\varphi :M\rightarrow\R^+$ as
$\tilde{g}:=\varphi^{-2}g$. Then the traceless Ricci part of the new
curvature is (see \cite{bes}, p. 59)
\[\widetilde{B}= {n-2\over\varphi}\left(\nabla^2\varphi
+{\triangle\varphi\over n}g\right) .\]  
Here $\triangle$ denotes the Laplacian with respect to $g$. From here we
can see that if $n>2$, the condition for $\tilde{g}$ to be again Einstein
is
\[\nabla^2\varphi +{\triangle\varphi\over n}g =0.\]
However, if $X:=\langle\dd\varphi ,\:\cdot\:\rangle$ is the dual vector
field then we can write by (\ref{dekompozicio}) that
\[\nabla^2\varphi ={1\over 2}\dd^2\varphi -{\delta (\dd\varphi
)\over n}g+{1\over 2}\left( L_Xg+{2\delta (\dd\varphi )\over n}g\right) =
-{\triangle\varphi\over n}g+{1\over 2}\left( L_Xg+{2\triangle\varphi\over 
n}g\right).\]
We have used $\dd^2=0$ and $\delta\dd =\triangle$ for functions. Therefore
we can conclude that $\varphi^{-2}g$ is Einstein if and only if
\[L_Xg+{2\triangle\varphi\over n}g=L_Xg-{2{\rm div}(X)\over n}g=0\]
i.e. $X$ is a conformal Killing field on $(M,g)$ obeying 
$X=\langle\dd\varphi ,\:\cdot\:\rangle$. The conformal Killing data
(\ref{adatok}) for this $X$ are the following:
\begin{equation}
(\dd\varphi ,\:\:\:\dd^2\varphi =0,\:\:\: -\triangle\varphi ,\:\:\:-\dd
(\triangle\varphi )).
\label{konkretadatok}
\end{equation}
Now we may argue as follows: the last equation of
(\ref{egyenletek}) implies that 
\[\nabla ( \dd (\triangle\varphi ))=0\]
over the Ricci-flat $(M, g)$. By virtue of the irreducibility of
$(M,g)$ this means that actually $\dd (\triangle\varphi )=0$ (cf.
e.g. \cite{bes}, p. 282, Th. 10.19) and hence $\triangle\varphi
=$const. over the whole $(M,g)$. Consequently, the second equation of
(\ref{egyenletek}) shows that for all $Y,Z,V$ we have
\[R(Y,Z,V, \dd\varphi )=0.\]
Taking into account again that $(M,g)$ is irreducible, there is a point
where $R_p$ is non-zero. Assume that the previous equality holds for all
$Y_p,Z_p,V_p$ but $\dd\varphi_p\not=0$. This is possible only if a
subspace, spanned by $\dd\varphi_p$ in $T^*_pM$, is invariant under the
action of the holonomy group. But this contradicts the irreducibility
assumption. Consequently $\dd\varphi_p=0$. Finally, the first equation of
(\ref{egyenletek}) yields that $\triangle\varphi (p)=0$ 
i.e. $\triangle\varphi =0$. Therefore we can conclude that in that
point all the conformal data (\ref{konkretadatok}) vanish implying $X=0$.
In other words $\varphi$ is a non-zero constant. $\Diamond$
\vspace{0.1in}

\noindent In light of this proposition, general Ricci-flat 
manifolds cannot be rescaled into Einstein manifolds in a non-trivial
way. Notice that the Taub-NUT space (see below) is
an irreducible Ricci-flat manifold. If this was not the
case, then, taking into account its simply connectedness and geodesic
completeness, it would split into a Riemannian product 
$(M_1\times M_2, g_1\times g_2)$ by virtue of the de Rham theorem
\cite{deR}. But it is easily checked that this is not the case. We just
remark that the same is true for the Euclidean Schwarzschild manifold. 

Consequently constructing instantons in this way is not very
productive. 

\section{Constructing new Taub-NUT instantons}
Therefore we turn our attention to the condition on the $F^-$ part of
the metric curvature in the special case of the Taub-NUT space. Consider 
the Taub-NUT metric
\begin{equation}
\dd s^2={r+m\over r-m}\dd r^2+4m^2{r-m\over r+m}(\dd\tau
+\cos\Theta\dd\phi
)^2+(r^2-m^2)(\dd\Theta^2+\sin^2\Theta\dd\phi^2),
\label{taub}
\end{equation}
\[m\leq r<\infty ,\:\:\:\:\:0\leq\tau <4\pi ,\:\:\:\:\:0\leq\phi <2\pi
,\:\:\:\:\:0\leq\Theta <\pi ,\]
where $m$ is a positive real number. If one introduces the  left-invariant
1-forms $\sigma_x$, $\sigma_y$, $\sigma_z$ on $S^3$ given by
\[\sigma_x=-\sin\tau\dd\Theta
+\cos\tau\sin\Theta\dd\phi ,\:\:\:\:\:\sigma_y=\cos\tau\dd\Theta
+\sin\tau\sin\Theta\dd\phi ,\:\:\:\:\:\sigma_z=\dd\tau +\cos\Theta\dd\phi
,\]
then the metric can be re-written in another useful form as
\[\dd s^2={r+m\over r-m}\dd r^2+(r^2-m^2)\left( \sigma_x^2+\sigma_y^2+
\left( {2m\over r+m} \right)^2 \sigma_z^2\right).\]
As it is well known (see e.g. \cite{pop}) this metric, despite its 
apparent singularity in the origin, extends analytically to a metric on
$\R^4$. Moreover it is Ricci-flat i.e. $B=0$ and $s=0$ as
well. Also notice that the Taub-NUT space is (anti)self-dual ($W^+=0$ or
$W^-=0$, depending on the orientation).

Our aim is to find metrics $\tilde{g}$ (as much as
possible), conformally equivalent to the original Taub-NUT metric $g$,
such that $\tilde{g}$'s are self-dual and have vanishing scalar curvature: 
in this case the metric instantons in $\tilde{g}$ provide (anti)self-dual
connections in the Taub-NUT case, as we have seen. Taking
into account that the $(3,1)$-Weyl tensor $W$ is invariant under
conformal rescalings i.e. $\widetilde{W}=W$, the condition
$\widetilde{W}^-+\tilde{s}/12 =0$ for the $\tilde{g}$'s settles down for
having vanishing scalar curvature $\tilde{s}=0$. Consider the rescaling
\[g\longmapsto\tilde{g}:=f^2g\]
where $f: \R^4\rightarrow\R^+$ is a positive function. In this case
the scalar curvature transforms as $s\mapsto\tilde{s}$ where $\tilde{s}$
satisfies (see \cite{bes}, pp. 58-59):
\[f^3\tilde{s}=6\triangle f +fs.\]
Taking into account that the Taub-NUT space is Ricci-flat, i.e. $s=0$, our
condition for the scaling function amounts to the simple condition
(cf. pp. 366-367 of \cite{boo-ble})
\begin{equation}
\triangle f =0
\label{laplace}
\end{equation}
i.e. it must be a harmonic function (with respect to the Taub-NUT
geometry). From (\ref{taub}) one easily calculates $\det
g=4m^2(r^2-m^2)^2\sin^2\Theta$ and by using the local expression
\[\triangle =\sum\limits_{i,j}{1\over\sqrt{\det g}}{\partial\over\partial
x^i}\left(\sqrt{\det g}\:g^{ij}{\partial\over\partial x^j}\right) ,\]
where $g^{ij}$ are the components of the inverse matrix, the Taub-NUT
Laplacian looks like
\[\triangle
=\left( {r+m\over 4m^2(r-m)}+{\cot^2\Theta\over r^2-m^2}\right)
{\partial^2\over\partial\tau^2}+{r-m\over
r+m}{\partial^2\over\partial r^2}+{2\over r+m}{\partial\over\partial
r}+\]
\[+{1\over
r^2-m^2}\left({\partial^2\over\partial\Theta^2}+\cot\Theta{\partial\over
\partial\Theta}\right) +{1\over 
(r^2-m^2)\sin^2\Theta}\left({\partial^2\over\partial\phi^2}-2\cos\Theta
{\partial^2\over\partial\tau\partial\phi}\right) .\]
The easiest thing is to seek for solutions depending only on the radial
coordinate $r$, in other words functions invariant under the full $U(2)$
symmetry of the Taub-NUT space. In this case the Laplace equation
(\ref{laplace}) cuts down to the ordinary differential equation
\[(r-m)f''+2f'=0\]
where prime denotes differentiation with respect to $r$. The general
solution to this equation is $f(r)=c_1+c_2(r-m)^{-1}$. Hence if we rescale
the original metric via 
\begin{equation}
f(r)=1+{\lambda\over r-m}
\label{skala}
\end{equation}
with $\lambda\geq 0$ then the metric (anti)instantons in $\tilde{g}=f^2g$
will be (anti)instantons in the original space. Although the metric
$\tilde{g}$ is singular in the origin, the resulting instantons will 
automatically appear in a gauge in which they are smooth everywhere.

Do not hesitate, let us construct these instantons explicitly in order to
determine their energy! Choose a global trivialization of the
spinor bundle $SM$ induced by the obvious orthonormal tetrad
\begin{equation}
\xi^0=\sqrt{{r+m\over r-m}}\:\dd
r,\:\:\:\:\:\xi^1=\sqrt{r^2-m^2}\:\sigma_x,\:\:\:\:\:\xi^2=
\sqrt{r^2-m^2}\:\sigma_y,\:\:\:\:\:\xi^3=2m\sqrt{{r-m\over
r+m}}\:\sigma_z.
\label{bazis} 
\end{equation}  
(Note that in four dimensions we have $SM\cong TM\otimes\C$). In this
gauge the Levi--Civit\'a connection of the re-scaled space is globally
represented by an $\so$-valued 1-form $\tilde{\omega}$ whose
components obey the Cartan equation
\[\dd\tilde{\xi}^i=-\tilde{\omega}^i_j\wedge\tilde{\xi}^j\]
where $\tilde{\xi}^i=f\xi^i$. Consequently we can write
\begin{equation}
\dd\xi^i+\dd (\log f)\wedge\xi^i=-\tilde{\omega}^i_j\wedge\xi^j.
\label{ujkonnexio}
\end{equation}
The basis (\ref{bazis}) shows that 
\[\dd (\log f)=\sqrt{{r-m\over r+m}}(\log f)'\xi^0.\]
Putting this and $\dd\xi^i=-\omega^i_j\wedge\xi^j$ into 
(\ref{ujkonnexio}) we get
\[\omega^i_j\wedge\xi^j+\sqrt{{r-m\over r+m}}(\log
f)'\xi^i\wedge\xi^0=\tilde{\omega}^i_j\wedge\xi^j.\]
Therefore, the components of the new connection form are given by
\begin{equation}
\tilde{\omega}^i_0=\omega^i_0+\sqrt{{r-m\over
r+m}}(\log f)'\xi^i,\:\:\:\:\:\tilde{\omega}^k_l=\omega^k_l
\label{konnexio}
\end{equation}
(here and only here $i,k,l$ run over 1,2,3). Now by using the original
connection (see \cite{egu-gil-han}, p. 351) and substituting
(\ref{skala}) into (\ref{konnexio}) we have the explicit form
\[\tilde{\omega}^1_0=\left({r\over r+m}-{\lambda\over
r-m+\lambda}\right)\sigma_x,\:\:\:\:\:\tilde{\omega}^3_2=-{m\over
r+m}\sigma_x,\]
\[\tilde{\omega}^2_0=\left({r\over r+m}-{\lambda\over
r-m+\lambda }\right)\sigma_y,\:\:\:\:\tilde{\omega}^1_3=-{m\over
r+m}\sigma_y,\]
\[\tilde{\omega}^3_0=\left({2m^2\over (r+m)^2}-{2m\lambda\over
(r-m+\lambda
)(r+m)}\right)\sigma_z,\:\:\:\:\:\tilde{\omega}^2_1=\left(
{2m^2\over (r+m)^2}-1\right)\sigma_z.\]
If we wish to project this connection to the self-dual part $A^-_\lambda$
we have to evoke the 't Hooft matrices which span the $\su^-\subset\so$
subalgebra:
\[\overline{\eta}_1=\pmatrix{ 0 & 1 & 0 & 0\cr -1 & 0 & 0 & 0\cr 0 & 0 & 0
& -1\cr 0 & 0 & 1 & 0\cr},\:\:\overline{\eta}_2=\pmatrix{0 & 0 & -1 & 0\cr
0 & 0 & 0 & -1\cr 1& 0 & 0 & 0\cr 0 & 1 & 0 & 0\cr},\:\:
\overline{\eta}_3=\pmatrix{0 & 0 & 0 & -1\cr 0 & 0 & 1 & 0\cr 0 & -1 & 0
& 0\cr 1 & 0 & 0 & 0\cr}.\]
The projected connection $A^-_\lambda$ in the gauge
(\ref{bazis}) is given by
\[A^-_\lambda ={1\over
4}\sum\limits_{a=1}^3\sum\limits_{i,j=0}^3\left(\overline{\eta}^i_{a,j}
\tilde{\omega}^i_j\right)\:\overline{\eta}_a.\]
The normalization factor $1/4$ comes from
$\overline{\eta}_a^{ij}\overline{\eta}_{a, ij}=4$ (summation is over
$i,j$ while $a$ is fixed). This eventually yields (by using the
identification $\su^-\cong{\rm Im}\HH$ via $(\overline{\eta}_1,
\overline{\eta}_2, \overline{\eta}_3)\mapsto (\ii ,\jj ,\kk )$ and the
straightforward gauge (\ref{bazis}))
\begin{equation}
A^-_\lambda =-{\ii\over 2}\Psi\sigma_x+
{\jj\over 2}\Psi\sigma_y +{\kk\over 2} \Phi\sigma_z 
\label{insztanton}
\end{equation}
where we have introduced the notations
\begin{equation}
\Phi (r):=1-{2m\lambda\over (r-m+\lambda )(r+m)},\:\:\:\:\:\Psi
(r):=1-{\lambda\over r-m+\lambda } .
\label{megoldas}
\end{equation}
These connections are self-dual with respect to the orientation
$\varepsilon_{\tau r\Theta\phi}=1$ and are non gauge-equivalent, 
as we will see in a moment. We also find that for $\lambda =0$ this
ansatz describes a singular flat connection. But this
isolated singularity can be removed by a gauge transformation, as
guaranteed by the Theorem of Removable Singularities by Uhlenbeck
\cite{uhl}. That is, $A_0^-$ is gauge-equivalent to
the trivial connection, since  $\R^4$ is simply connected. Otherwise
$A^-_\lambda$ is regular in the origin since $\Phi (m)=0$ and
$\Psi (m)=0$ if $\lambda >0$. In particular, for $\lambda =2m$ it reduces
to the Pope--Yuille instanton \cite{pop-yui}. Last but not least, if
$\lambda =\infty$, that is when the rescaling function is $1/(r-m)$, the
solution (\ref{insztanton}) takes the form
\[A^-_\infty ={\kk\over 2}{r-m\over r+m}\sigma_z,\]
which coincides with the Abelian instanton found
by Eguchi--Hanson \cite{egu-han} and reinvented in
mathematical terms by Gibbons \cite{gib} (see also \cite{pop}).

Now we calculate the curvature $F^-_\lambda =\dd A^-_\lambda +{1\over
2}[A^-_\lambda ,A^-_\lambda ]=\dd A^-_\lambda
+A^-_\lambda\wedge A^-_\lambda$. Taking into account the identities 
$\dd\sigma_x=\sigma_y\wedge\sigma_z$, etc. (the indices $x,y,z$ are
permuted cyclically), a straightforward calculation yields the shape
\[F^-_\lambda ={\ii\over 2}\left( -\Psi '\dd r\wedge\sigma_x -(\Psi
-\Psi\Phi )\sigma_y\wedge\sigma_z\right) +\]
\begin{equation}
+{\jj\over 2}\left(\Psi '\dd r\wedge\sigma_y-(\Psi
-\Psi\Phi )\sigma_x\wedge\sigma_z\right) + 
\label{gorbulet}
\end{equation}
\[+{\kk\over 2}\left (\Phi '\dd r\wedge \sigma_z+(\Phi -\Psi^2 
)\sigma_x\wedge\sigma_y\right) .\]
If $A,B\in\su$ whose images are $x,y\in{\rm Im}\HH$ under the above
identification, then, as easily checked, the Killing form $-\tr (AB)$ is
given by $2{\rm Re}(x\overline{y})$. Bearing this in mind, the energy-density 
4-form is  
\[\vert F^-_\lambda\vert^2_g =-\tr (F^-_\lambda\wedge F^-_\lambda
)=\left(\Phi '\Phi +2\Psi '\Psi -2\Psi '\Psi\Phi-\Phi '\Psi^2\right)\dd
r\wedge\sigma_x\wedge\sigma_y\wedge\sigma_z=\]
\begin{equation}
=\left(\Phi '\Phi +2\Psi '\Psi -2\Psi '\Psi\Phi-\Phi
'\Psi^2\right) 
\sin\Theta\dd\tau\wedge\dd r\wedge\dd\Theta\wedge\dd\phi .
\label{energiasuruseg}
\end{equation}
Using the parameterization (\ref{taub}), the energy is the integral
\[\Vert F^-_\lambda\Vert^2_{L^2(\R^4)}=\]
\[={1\over  
8\pi^2}\int\limits_0^{2\pi}\int\limits_0^{\pi}\int\limits_m^\infty\int
\limits_0^{4\pi}\left(\Phi '(r)\Phi (r) +2\Psi '(r)\Psi (r)
-2\Psi '(r)\Psi (r)\Phi (r)-\Phi '(r)\Psi^2(r)\right)\sin\Theta\dd\tau\dd
r\dd\Theta\dd\phi .\]
By performing the only non-trivial integral 
\[\int\limits_m^\infty\left(\Phi '(r)\Phi (r) +2\Psi '(r)\Psi
(r) -2\Psi '(r)\Psi (r)\Phi (r)-\Phi '(r)\Psi^2(r)\right)\dd r=
\left\{ \begin{array}{ll}
                  0 & \mbox{if $\lambda =0$} \\
                 1/2 & \mbox{ if $\lambda >0$}
        \end{array}
\right.\]
we eventually find the energy to be 
\[\Vert F^-_\lambda\Vert^2_{L^2(\R^4)}=
\left\{\begin{array}{ll}
                  0 & \mbox{if $\lambda =0$} \\
                  1 & \mbox{ if $\lambda >0$.}
       \end{array}
\right.\]
To conclude this section, we show that the connections $A_\lambda^-$ just
constructed do indeed satisfy the self-duality equations and are
not gauge equivalent. 

Using the orthonormal tetrad (\ref{bazis}), it follows that the
Hodge-operator on 2-forms is given as follows: 
\[*(\dd r\wedge \sigma_x)= *\left({1\over
r+m}\xi^0\wedge\xi^1   \right)={1\over r+m}\xi^2\wedge\xi^3 =
{2m(r-m)\over r+m}\sigma_y\wedge \sigma_z,\] 
\[*(\dd r\wedge \sigma_y)= *\left({1\over
r+m}\xi^0\wedge\xi^2\right)=-{1\over r+m}\xi^1\wedge\xi^3 =-{2m(r-m)\over
r+m}\sigma_x\wedge \sigma_z,\]
\[*(\dd r\wedge \sigma_z)= *\left({1\over 2m}\xi^0\wedge\xi^3\right)=
{1\over 2m}\xi^1 \wedge\xi^2 = {r^2-m^2\over 2m} \sigma_x\wedge \sigma_y. 
\]
Applying this to the original curvature
expression (\ref{gorbulet}) we get 
\[*F^-_\lambda ={\ii\over 2}\left(-{2m(r-m)\over
r+m}\Psi '\sigma_y\wedge\sigma_z-{r+m\over 2m(r-m)}(\Psi -\Psi\Phi )\dd
r\wedge\sigma_x\right)+   \]
\[+{\jj\over 2}\left( -{2m(r-m)\over r+m}\Psi '\sigma_x\wedge\sigma_z+
{r+m\over 2m(r-m)}(\Psi -\Psi\Phi )\dd r\wedge\sigma_y\right)+\]
\[+{\kk\over 2}\left({r^2-m^2\over 2m}\Phi '\sigma_x\wedge\sigma_y+{2m\over
r^2-m^2}(\Phi -\Psi^2)\dd r\wedge \sigma_z\right) .\]
Therefore comparing the original curvature (\ref{gorbulet}) with the above
expression term by term, the self-duality equations reduce to the two
equations 
\[\Psi '= {r+m\over 2m(r-m)}(\Psi -\Psi\Phi ),\]
\[\Phi '= {2m\over r^2-m^2} (\Phi -\Psi^2). \]
One checks by hand that the previously constructed functions obey these
equations. We note that the above equations were previously obtained in
\cite{kim-yoo}. 

Also notice that our solutions are non-gauge
equivalent. The simplest way to see this is by looking at the
energy-density, which is a gauge independent $4$-form on the space. 
Putting the functions (\ref{megoldas}) into the
energy-density 4-form (\ref{energiasuruseg}) then for small $\lambda$ we
get
\[\vert F^-_\lambda\vert^2_g\sim {4\lambda^2m(3r^3-r^2m-rm^2-m^3)\over
(r-m+\lambda )^4(r+m)^3}\dd r\wedge\sigma_x\wedge\sigma_y\wedge\sigma_z.\]
Consequently the energy-density approaches a Dirac-delta 
concentrated at the origin $r=m$ of $\R^4$ as $\lambda\rightarrow
0$. This shows that the connections with different $\lambda$'s are
non-gauge equivalent because their curvatures are different. This
observation also provides a natural interpretation of the parameter
$\lambda$, namely it measures the ``concentration'' of the instanton 
around the origin of $\R^4$. The limit connection as $\lambda\rightarrow
0$ is an ``ideal'', Dirac-delta-type instanton, concentrated at the NUT. 
This ideal connection compactifies our half line of solutions into a closed
segment. 

\section{Concluding remarks}

In this paper we constructed a one-parameter family of new Taub-NUT instantons
of unit energy
depending on a parameter $\lambda\in (0,\infty]$. When $\lambda=2m$ we
recovered the previously known Pope--Yuille instanton and when
$\lambda=\infty$ we obtained a reducible Maxwell instanton corresponding
to the Eguchi--Hanson--Gibbons $L^2$ harmonic $2$-form. 

Our hope is that there exist nice moduli spaces of Yang--Mills instantons
on Taub-NUT
space like for the flat $\R^4$ (cf. \cite{ati-hit-sin}). Unfortunately at
present even the Atiyah--Singer index theorem is not fully understood on
manifolds with ends like Taub-NUT (however cf. \cite{pop},
\cite{nye-sin} and \cite{vai}). So the infinitesimal description of the
moduli space lacks some 
fundamental analysis. Nevertheless we hope that the half-line of solutions we
found will actually turn out to be the full moduli space of unit
charge $SU(2)$ Yang--Mills instantons on
Taub-NUT space. To go further one might conjecture that the moduli space
of {\em framed} $SU(2)$ instantons of unit charge on Taub-NUT space will
be the Taub-NUT space itself, and the half-line we found will arise as
Taub-NUT space modulo the isometric action of $SU(2)\subset U(2)$ (moving
the frame). In this description the NUT would correspond to the reducible
instanton, the end-point of the half-line. 

Another intriguing possibility is to analyse what happens if we consider
our $U(1)$-invariant Yang--Mills solutions as solutions of the Bogomolny
equations on $\R^3=\R^4/U(1)$ (c.f. \cite{kro}). We hope to return to
this problem in the near future.

Finally we mention that the analogous constructions on the
Gibbons--Hawking multi-centered spaces will appear elsewhere
\cite{ete-hau3}. There we will see a clearer understanding of the role of
harmonic functions and their singularities in the problem. Also we will
find there nice geometrical constructions of the $L^2$-harmonic forms on
these multi-centered spaces, whose existence is proven by different
means in \cite{hau-maz}.

\end{document}